\documentclass[
  prx,
  superscriptaddress,
  aps,
  amsmath,
  amssymb,
  floatfix,
  reprint,
  raggedbottom,
  longbibliography
]{revtex4-2}

\usepackage{graphicx}
\usepackage{hyperref}
\usepackage{xcolor}
\usepackage{algorithm}
\usepackage{algpseudocode}

\hypersetup{
    colorlinks=true,
    linkcolor=red,
    citecolor=blue,
    urlcolor=black
}

\begin{document}

\title{Restoring Sparsity in Potts Machines via Mean-Field Constraints}

\author{Kevin Callahan-Coray}\email{kevincallahan-coray@ucsb.edu}\thanks{Corresponding Author}\affiliation{Department of Electrical and Computer Engineering, 
University of California, Santa Barbara, Santa Barbara, CA 93106, USA}
\author{Kyle Lee}
\affiliation{Department of Electrical and Computer Engineering, 
University of California, Santa Barbara, Santa Barbara, CA 93106, USA}
\author{Kyle Jiang}
\affiliation{Department of Electrical and Computer Engineering, 
University of California, Santa Barbara, Santa Barbara, CA 93106, USA}
\author{Kerem Y. Çamsarı}
\affiliation{Department of Electrical and Computer Engineering, 
University of California, Santa Barbara, Santa Barbara, CA 93106, USA}


\begin{abstract}
Ising machines and related probabilistic hardware have emerged as promising platforms for NP-hard optimization and sampling. However, many practical problems involve constraints that induce dense or all-to-all couplings, undermining scalability and hardware efficiency. We address this constraint-induced density through two complementary approaches. First, we introduce a hardware-aware native formulation for multi-state probabilistic digits (p-dits) that avoids the locally dense intra-variable couplings required by binary Ising encodings. We validate p-dit dynamics by reproducing known critical behavior of the 2D Potts model. Second, we propose mean-field constraints (MFC), a hybrid scheme that replaces dense pairwise constraint couplings with dynamically updated single-node biases. Applied to balanced graph partitioning, MFC achieves solution quality comparable to exact all-to-all constraint formulations while dramatically reducing graph density. Finally, we demonstrate the practical impact of restored sparsity through an FPGA implementation. In comparisons using FPGA kernel time and CPU solver-loop time, and excluding the current prototype’s host-device schedule-transfer overhead, the FPGA reaches the 50\% success threshold more than an order of magnitude faster than the CPU probabilistic solvers and more than two orders of magnitude faster than the Tabu Ising baseline. Together, these results outline a pathway for scaling constrained optimization on probabilistic hardware.
\end{abstract}

\maketitle

\section{Introduction}
\label{sec:introduction} 

Ising machines and related probabilistic computing architectures have emerged as promising platforms not only for accelerating NP-hard optimization but also for sampling tasks relevant to machine learning and as components in hybrid classical-quantum workflows~\cite{mohseni2022isingmachineshardwaresolvers,mohseni2024quantum,YFu_1986}. A key factor underlying their reported scalability is the sparsity of the interaction graph, which enables localized updates, low communication overhead, and efficient implementation across digital~\cite{Aadit_2022,Sajeeb_2025}, analog~\cite{Chou2019}, and mixed-signal~\cite{Singh2024} hardware platforms. As a result, much of the existing work on Ising machines has focused on unconstrained or weakly constrained optimization problems, where sparsity is preserved and large systems can be simulated or realized efficiently~\cite{Aadit_2022}. 

In contrast, many relevant optimization problems are dominated by constraints that fundamentally alter this sparsity structure. Constraints such as balancing, cardinality, exclusivity, and resource conservation often introduce dense or all-to-all interactions, even when the underlying objective function is sparse~\cite{Lucas_2014}. Enforcing such constraints through penalty terms or auxiliary variables leads to interaction graphs whose density scales poorly with problem size~\cite{Sajeeb_2025}, eroding parallelism and substantially increasing computational and hardware costs. As a result, constraint enforcement emerges as a primary obstacle to scaling Ising machines for real-world applications.

This work addresses constraint-induced density by introducing two complementary mechanisms that restore sparsity in constrained probabilistic optimization. First, we propose a hardware efficient probabilistic digit (p-dit) \cite{duffee2025pdit}, a multi-state generalization of probabilistic bits (p-bits)~\cite{Kaiser_2021} that absorbs local constraints directly into the node state space. By embedding mutually exclusive configurations into a single multi-state variable, p-dits eliminate dense intra-variable penalty couplings while preserving the exact local energy structure. Second, we introduce mean-field constraints (MFC), a hybrid probabilistic-classical approach in which global constraints are enforced approximately through dynamically updated bias signals. This decouples global constraint enforcement from local stochastic updates, replacing dense pairwise interactions with a shared, slowly varying mean field.

We demonstrate the effectiveness of these approaches using the balanced graph partitioning problem, a canonical example of constrained optimization that combines a sparse objective with a global balancing constraint. Using this problem as a test case, we show that mean-field constraints achieve solution quality comparable to strictly constrained formulations while dramatically reducing effective graph density. We benchmark against a Tabu Ising baseline and validate the practical impact of restored sparsity through an FPGA implementation, showing that constrained optimization problems can recover the parallelism and performance of sparse Ising machines once constraint-induced density is removed.

\begin{figure*}[t]
    \centering
    \includegraphics[width=.9\linewidth]{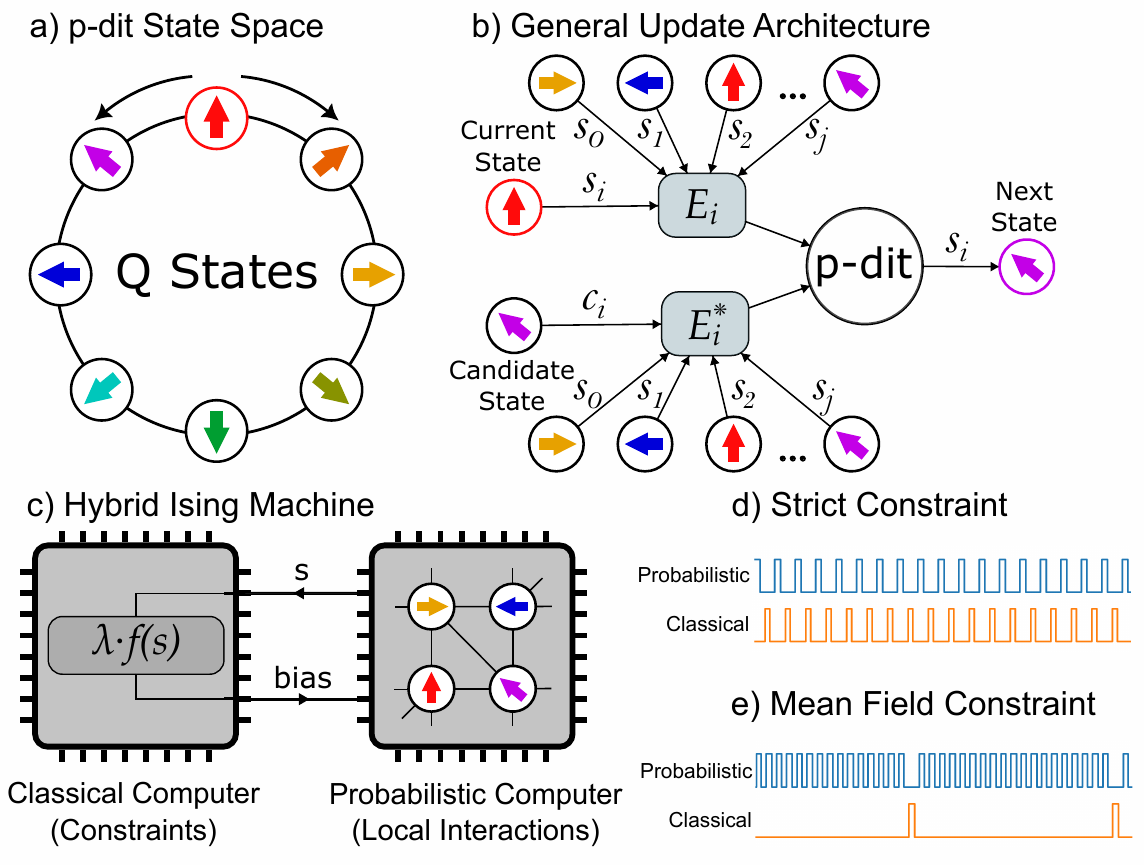}
   \caption{\textbf{Overview of the hybrid probabilistic-classical framework.}
(a)~p-dit state space: A p-dit occupies one of $Q$ discrete states arranged on a ring and proposes transitions only to neighboring states.
(b)~Update rule: The p-dit computes the local energy of its current state ($E_i$) and a candidate neighbor ($E_i^*$), then transitions probabilistically based on the difference.
(c)~Hybrid architecture: The probabilistic subsystem handles local interactions. A classical process computes global constraints and broadcasts a bias signal. Both may reside on the same physical fabric, the separation is conceptual.
(d)~Strict constraints require the classical process to update at every p-dit flip, imposing high synchronization overhead.
(e)~Mean-field constraints approximate the constraint as a slowly varying field, allowing updates only once per sweep and reducing classical workload while guiding the system toward feasibility.}

    \label{fig:overview}
\end{figure*}

\begin{figure*}[t]
    \centering
    \includegraphics[width=1.0\linewidth]{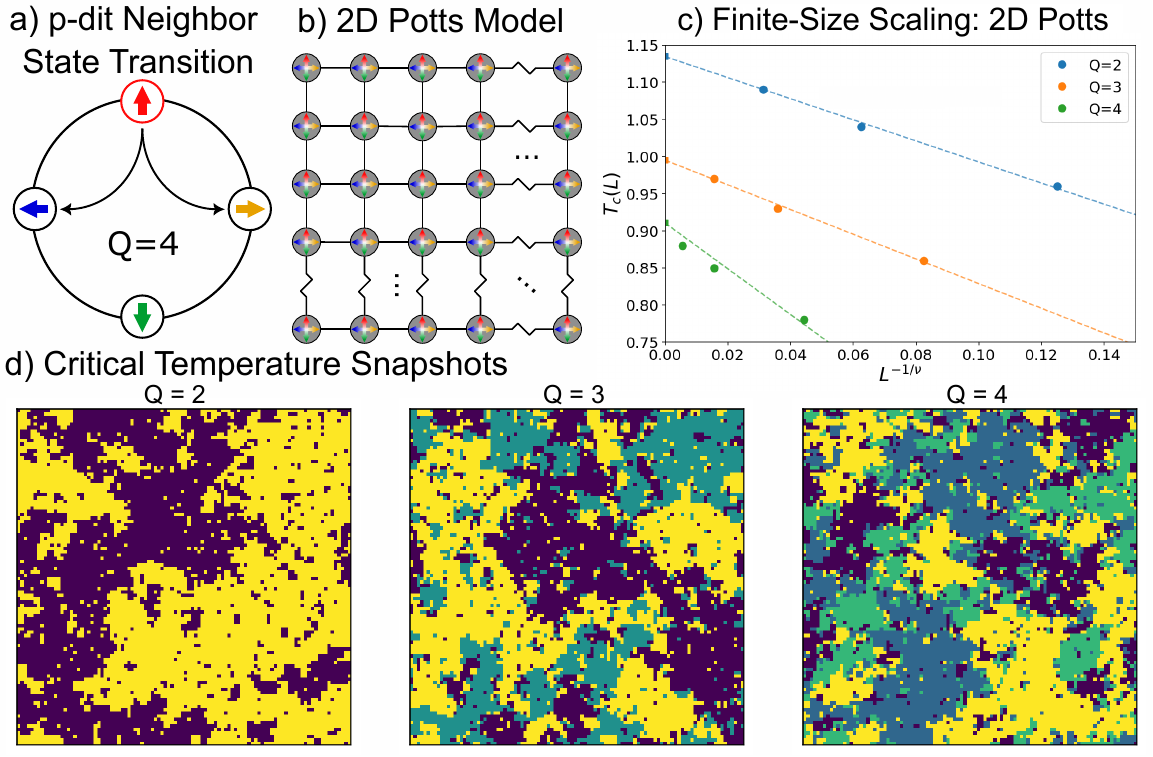}
  \caption{\textbf{Verification of p-dit dynamics using the 2D Potts model.}
(a)~Four-state p-dit showing discrete angular states and allowed transitions between nearest neighbors.
(b)~2D Potts lattice used for validation: $L \times L$ square lattice with nearest-neighbor interactions.
(c)~Finite-size scaling: Measured critical temperatures $T_c(L)$ for $Q=2,3,4$ follow $T_c(L)\approx T_c(\infty)+aL^{-1/\nu}$~\cite{PhysRevLett.28.1516,PhysRevB.30.322,PhysRevB.30.1477}, consistent with known 2D Potts behavior.
(d)~Configurations at criticality: Representative $L=100$ configurations for $Q=2,3,4$ at the fitted critical temperatures.}

    \label{fig:pdit}
\end{figure*}

\section{Multi-state probabilistic digits (p-dits)}
\label{sec:pdit}

The Potts model generalizes the Ising model, where binary spins $m_i \in \{-1,+1\}$ evolve to minimize $E = -\sum_{i<j} J_{ij} m_i m_j - \sum_i h_i m_i$, by allowing each variable to occupy one of $Q$ discrete states, $s_i \in \{0,1,\ldots,Q-1\}$~\cite{wureview1982}. The most general first-order Potts Hamiltonian is
\begin{equation}
    E = -\sum_{i<j} J_{ij}(s_i,s_j) - \sum_i h_i(s_i),
\end{equation}
where $J_{ij}(s_i,s_j)$ defines the interaction energy between states of nodes $i$ and $j$, and $h_i(s_i)$ is a state-dependent bias.

Many optimization problems involve variables with more than two states (e.g., coloring, partitioning) that map directly onto Potts spins. Encoding such problems in binary requires multiple coupled bits per variable, expanding the state space to include infeasible configurations that must be suppressed through penalties~\cite{Lucas_2014}. Native Potts encodings avoid this overhead entirely, as  recent work has shown that eliminating infeasible states alone can accelerate convergence by orders of magnitude~\cite{iyer2025efficient}. Potts-based hardware has been demonstrated on CMOS-compatible architectures~\cite{whitehead2023cmos}, coupled ring-oscillator machines~\cite{gonul2024potts}, coherent optical implementations~\cite{inoue2022coherent}, and p-dit formulations~\cite{duffee2025pdit}. 

In this work, we introduce a hardware-aware p-dit update rule based on local energy comparisons between neighboring Potts states, applicable to general Potts Hamiltonians and validated through finite-size scaling of the 2D Potts model. Fig.~\ref{fig:overview}a and~\ref{fig:overview}b illustrate the p-dit state space and update structure. For global constraints that cannot be absorbed locally, we introduce mean-field constraints (Figs.~\ref{fig:overview}c--e). Throughout this work, formulations in which such constraints are evaluated exactly are referred to as \emph{strictly constrained} and serve as baselines.

The p-dit is directly inspired by the probabilistic bit (p-bit)~\cite{Kaiser_2021,Camsari_2017} and can be viewed as its multi-state generalization. In the p-bit framework, each binary variable stochastically updates based on a local interaction field $I_i = \sum_j J_{ij} m_j + h_i$, which corresponds to the energy difference between the two spin states. The probability of being $+1$ is
\begin{equation}
    p[m_i = +1] = \sigma(2\beta I_i),
    \label{eq:pbit_prob}
\end{equation}
where $\sigma(\cdot)$ is the sigmoid function, and $\beta$ is the inverse temperature. 

The p-dit extends this to multi-state variables by replacing binary flips with pairwise comparisons. Each node maintains a current state $s_i$ and evaluates a candidate state $c_i$. The energy difference is
\begin{equation}
    E(s_i \rightarrow c_i) =
    \sum_j J_{ij}(s_i \rightarrow c_i, s_j)
    + h_i(s_i \rightarrow c_i),
    \label{eq:delta_e_potts}
\end{equation}
where functions are expressed in a shorthand form which represents the change in value from altering an input, i.e. $f(x \rightarrow y, ...) \equiv f(x, ...) - f(y,...)$. Only the local energy difference need be evaluated. The candidate is accepted with probability
\begin{equation}
    p[s_i = c_i] = \sigma\!\left(\beta E(s_i \rightarrow c_i)\right),
    \label{eq:pdit_prob}
\end{equation}
where $E(s_i \rightarrow c_i)$ is the energy of the current state minus that of the candidate, so candidates with lower energy than the current state are accepted with probability greater than one half.

Each update consists of candidate selection followed by stochastic acceptance. We employ nearest-neighbor selection on a circular Potts state space (Fig.~\ref{fig:pdit}a); the candidate is chosen uniformly from the two adjacent states. This ensures symmetric proposals while minimizing control complexity. Restricting candidates to neighbors slows mixing but greatly simplifies hardware. Crucially, equilibrium correctness is preserved; the dynamics satisfy detailed balance and converge to the Boltzmann distribution (Sec.~\ref{subsec:detailed_balance}).

To validate statistical correctness, we performed finite-size scaling of the 2D ferromagnetic Potts model on $L \times L$ lattices for $L \in \{8,16,32\}$ and $Q \in \{2,3,4\}$ (Fig.~\ref{fig:pdit}b). The extracted critical temperatures follow $T_c(L) \approx T_c(\infty) + aL^{-1/\nu}$~\cite{PhysRevLett.28.1516,PhysRevB.30.322,PhysRevB.30.1477}, with strong agreement to exact values (Fig.~\ref{fig:pdit}c). Representative configurations at criticality exhibit the expected domain structure (Fig.~\ref{fig:pdit}d). These results establish the p-dit as a statistically correct, hardware-efficient primitive for multi-state probabilistic computing. By embedding constraints into node state spaces and using sequential two-state comparisons, p-dits restore sparsity without sacrificing equilibrium correctness.

While p-dits eliminate constraint-induced density at the node level, many problems impose global constraints that cannot be encoded locally. Addressing these without reintroducing dense couplings motivates the mean-field approach introduced next.

\begin{figure*}[t]
    \centering
    \includegraphics[width=0.9\linewidth]{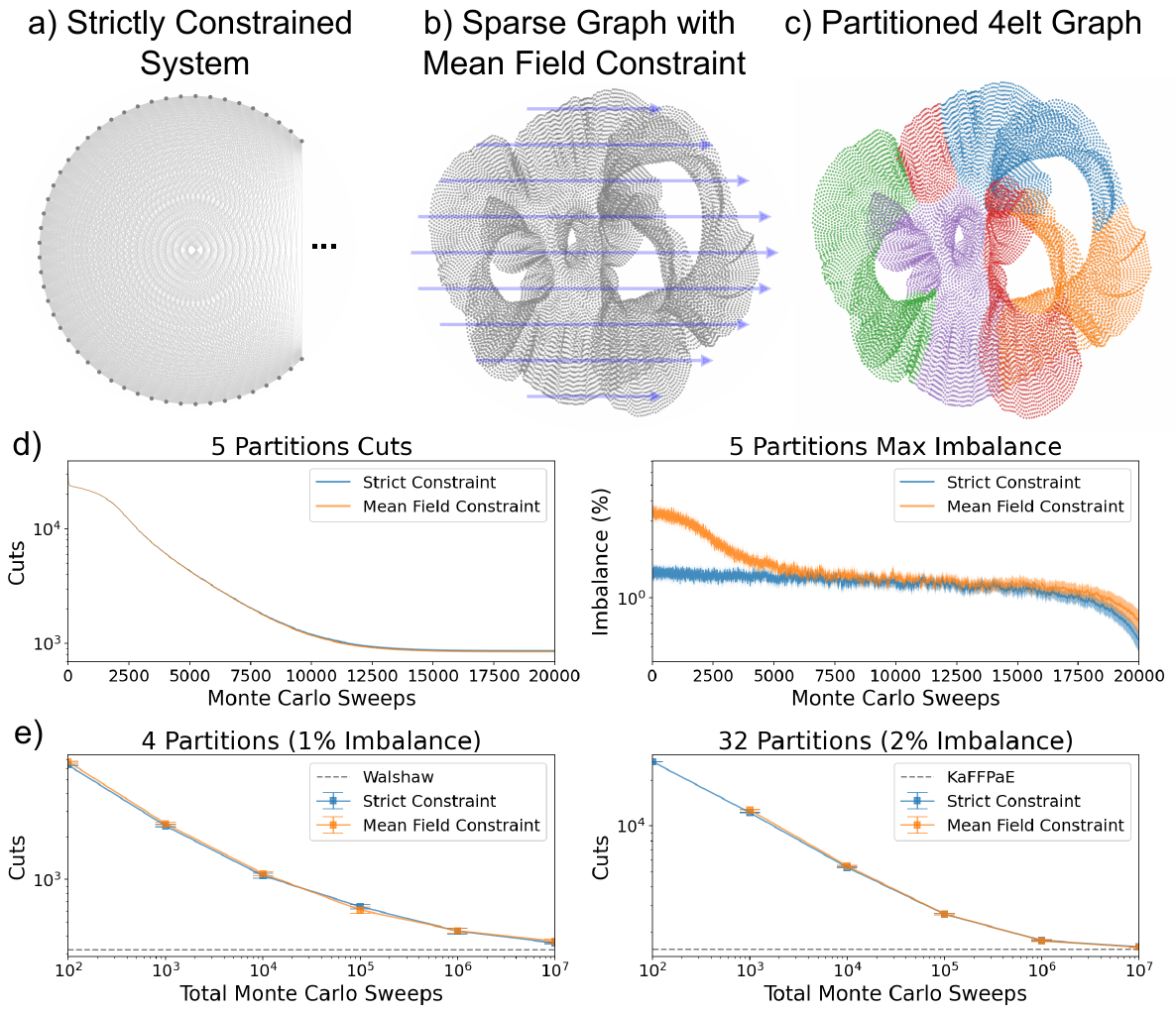}
   \caption{\textbf{Mean-field constraints for balanced minimum-cut partitioning.}
(a)~Strictly constrained system: The \texttt{4elt} mesh encoded with all-to-all constraint couplings.
(b)~Mean-field formulation: The same problem with dense constraint couplings replaced by a global bias signal (blue arrows).
(c)~Example result: A 5-way partition of \texttt{4elt} after $10^{6}$ Monte Carlo sweeps.
(d)~Dynamics: Evolution of cut size and partition imbalance for strict vs.\ mean-field constraints, averaged over 100 simulated annealing trials.
(e)~Benchmarking: Solution quality for 4-way (1\% maximum imbalance) and 32-way (2\% maximum imbalance) partitions. Dashed lines show reference solutions (Walshaw for 4-way; KaFFPaE for 32-way).}
    \label{fig:mfc}
\end{figure*}

\section{Restoring sparsity using Mean-Field Constraints}
\label{sec:mfc}

Many optimization problems impose constraints that depend on global properties of the system~\cite{Lucas_2014}. Because these constraints act collectively, they cannot be absorbed into local state spaces. When enforced strictly, such global constraints introduce dense or all-to-all interactions, destroying sparsity and limiting parallelism~\cite{Aadit_2022,Sajeeb_2025}.

Mean-field constraints provide a relaxed alternative; rather than coupling nodes through pairwise constraint terms, the constraint influence is approximated by a shared bias updated from the aggregate system state. This replaces constraint-induced density with a single global signal while retaining a guiding force toward feasibility (Figs.~\ref{fig:mfc}a--b). Classical mean-field approximations are well established in statistical physics but can exhibit oscillatory or unstable behavior in strongly constrained systems~\cite{basak2016universalitymeanfieldpottsmodel,jain2018meanfieldapproximationinformationinequalities}. Stabilizing these dynamics for hardware implementation requires additional care.
We adopt a hybrid architecture combining probabilistic and classical computation (Fig.~\ref{fig:overview}c). The probabilistic subsystem performs local stochastic updates on the sparse interaction graph. A classical subsystem monitors the global state, evaluates constraint violations, and broadcasts a mean-field bias to all nodes. The bias is updated once per Monte Carlo sweep rather than at every node update. While this decoupling restores sparsity, directly applying a sweep-level bias can cause oscillations as the constraint response overcorrects.

To address this, we interpret the system through a control-theoretic lens: the probabilistic subsystem is the plant, and the classical subsystem is a feedback controller~\cite{Astrom_Murray_2010}. The constraint violation serves as an error signal, and the mean-field bias is the control input. A simple low-pass filter stabilizes the dynamics:
\begin{equation}
    \hat{\varepsilon}_n = \alpha \varepsilon_n + (1-\alpha)\hat{\varepsilon}_{n-1},
    \label{eq:low_pass}
\end{equation}
\begin{equation}
    h_{MFC} = \lambda f(\hat{\varepsilon}_n),
    \label{eq:mfc_bias}
\end{equation}
where $\varepsilon$ is the instantaneous constraint error, $\lambda$ controls constraint strength, and $\alpha$ sets the filter rate. The error $\varepsilon_n$ is evaluated at the end of sweep $n$, and the resulting bias is applied during sweep $n+1$, so the probabilistic and classical subsystems operate sequentially within each sweep. This suppresses rapid fluctuations while maintaining a consistent bias toward feasibility.
 
The mechanism generalizes readily. Multiple independent mean-field constraints can be enforced simultaneously, and locally dense constraints on node clusters can also be approximated using shared biases. For p-dits, the scalar bias extends to a vector over the state space~\cite{wureview1982,basak2016universalitymeanfieldpottsmodel}. Each component corresponds to the energy bias for a particular state, and the contribution enters additively into the local energy difference during updates.

\begin{figure*}[t]
    \centering
    \includegraphics[width=1.0\linewidth]{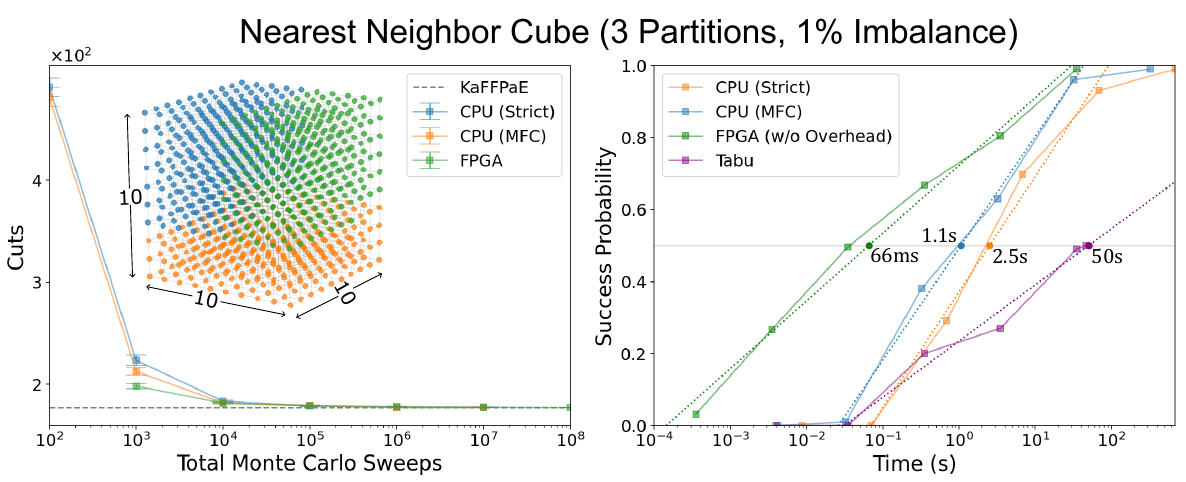}
    \caption{\textbf{Hardware performance for balanced min-cut partitioning.}
Left: cut-quality convergence for CPU-based strict constraints, CPU-based MFC, and FPGA-based MFC on a $10\times10\times10$ cube partitioned into three groups (1\% maximum imbalance). The approaches show slightly different dynamics due to the FPGA's precision and annealing approximations, but all converge toward the KaFFPaE reference.
Right: time-to-solution, the probability of reaching the KaFFPaE reference cut versus the time given to each solver. D-Wave's Tabu solver~\cite{dwave-tabu} serves as a heuristic Ising baseline on CPU. The FPGA curve is plotted without host-device overhead, since the beta schedule was uploaded at runtime; a realistic implementation would hold the schedule on chip and remove this runtime schedule-transfer overhead. Dotted lines show a linear regression in log-time used to locate the 50\% crossing.}

    \label{fig:hardware_advantage}
\end{figure*}

\section{Graph Partitioning}
\label{sec:graph_partitioning}

The graph partitioning problem seeks to divide vertices into multiple communities while minimizing edges crossing between them and maintaining approximately equal-sized partitions~\cite{Karypis_1998,buluc2015recentadvancesgraphpartitioning}. This problem provides a representative test case for constrained optimization, where a sparse objective (cut minimization) is combined with a global constraint (balance).

In the Potts formulation, the Hamiltonian is
\begin{equation}
    H = H_{\text{min}} + \lambda H_{\text{bal}},
\end{equation}
where $H_{\text{min}}$ encodes the cut objective and $H_{\text{bal}}$ enforces balance, with an adjustable constraint parameter $\lambda$. The minimum-cut term rewards neighboring vertices in the same partition:
\begin{equation}
    H_{\text{min}} = -\sum_{i<j} J_{ij}\,\delta(s_i,s_j),
\end{equation}
where $J_{ij}$ is the adjacency matrix and $\delta(\cdot, \cdot)$ is the Kronecker delta. This term is sparse whenever the underlying graph is sparse. Without a balancing constraint, minimizing $H_{\text{min}}$ alone yields a trivial solution where all vertices collapse into one partition. The balancing term penalizes deviations from equal sizes:
\begin{equation}
    H_{\text{bal}} = \sum_{k=0}^{Q-1} \left( \sum_i \delta(s_i,k) - \frac{N}{Q} \right)^2,
    \label{eq:hbal}
\end{equation}
where $Q$ is the number of partitions and $N$ the number of vertices. This term couples all vertices in each partition, inducing dense all-to-all interactions~\cite{Lucas_2014}.

Using p-dits, partition assignments are encoded directly as multi-state variables. Strict enforcement of $H_{\mathrm{bal}}$ preserves correctness but requires sequential updates, negating parallelism. Under mean-field relaxation, the dense balancing edges are replaced by a bias vector reflecting global imbalance. The constraint violation for each partition is
\begin{equation}
\varepsilon_{n,k} = \sum_i \delta(s_i, k) - \frac{N}{Q},
\end{equation}
filtered according to Eq.~\eqref{eq:low_pass}. Because p-dit transitions are state dependent, this bias pushes the system toward balance, and every pair of states involved in a candidate transition feels the same force. The general bias in Eq.~\eqref{eq:mfc_bias} applies a problem-specific function $f$ to the filtered error. For graph partitioning, the relevant quantity is the change in the balance penalty when a node moves from one partition to another. Moving a node from partition $a$ to partition $b$ changes the penalty in Eq.~\eqref{eq:hbal} by
\begin{equation}
    \Delta H_{\mathrm{bal}} = [(\hat{\varepsilon}_a - 1)^2 + (\hat{\varepsilon}_b + 1)^2] - [\hat{\varepsilon}_a^2 + \hat{\varepsilon}_b^2]
\end{equation}
In the $E(\text{current}) - E(\text{candidate})$ convention of Eq.~\eqref{eq:pdit_prob}, the bias is therefore
\begin{equation}
    h_{\mathrm{MFC}} = 2\lambda\big(\hat{\varepsilon}_a - \hat{\varepsilon}_b - 1\big).
    \label{eq:state_delta}
\end{equation}
All implementations follow this same form, with the factor of two absorbed into $\lambda$ for simplicity.

To evaluate MFC effectiveness, we compared strict and mean-field enforcement on the \texttt{4elt} benchmark graph~\cite{Walshaw_Archive} (Fig.~\ref{fig:mfc}c). Simulated annealing with 100 independent trials showed comparable cut quality between methods, with MFC exhibiting initially larger imbalance that rapidly converges (Fig.~\ref{fig:mfc}d). Quantitatively, the two methods track each other closely: MFC and strict enforcement differ by at most 2.25\% in mean cut across the entire anneal and by 1.02\% at convergence. Both methods improve with additional sweeps, approaching reference solutions from state-of-the-art solvers~\cite{Walshaw_Archive,sanders2011distributedevolutionarygraphpartitioning} (Fig.~\ref{fig:mfc}e). At the end of the sweep ranges in Fig.~\ref{fig:mfc}e, MFC is within 2.8\% of strict for 4-way partitioning and within 0.7\% for 32-way, with MFC giving the slightly lower cut in the 32-way case, so the mean-field relaxation does not sacrifice solution quality. For 32-way partitioning, an early Monte Carlo sweep data point for MFC is missing because the 2\% imbalance threshold could not be maintained, highlighting that MFC requires more sweeps for its low-pass filtering to become effective.

\section{Hardware Implementation and Performance}
\label{sec:hardware}

Mean-field constraints trade exact Boltzmann dynamics~\cite{jain2018meanfieldapproximationinformationinequalities} for scalable execution on parallel hardware. The preceding results were obtained on CPU. Here we demonstrate the practical advantage on an FPGA.
 
We implement a $10\times10\times10$ cubic graph partitioned into three balanced groups (Fig.~\ref{fig:hardware_advantage}). Both CPU and FPGA use identical p-dit dynamics, so differences arise primarily from the constraint mechanism and precision approximations in hardware. Architecture details and parameters appear in Sec.~\ref{sec:methods_fpga} and \ref{sec:anneal}.
 
Fig.~\ref{fig:hardware_advantage} compares convergence and solver runtime for CPU-based strict constraints, CPU-based MFC, FPGA-based MFC, and a Tabu baseline. All implementations converge toward similar cut values, consistent with KaFFPaE reference solutions~\cite{sanders2011distributedevolutionarygraphpartitioning}. The convergence dynamics differ slightly because the FPGA uses fixed-point arithmetic and a staircase annealing schedule, but the final cut values match. The Tabu solver~\cite{dwave-tabu} serves as the heuristic Ising baseline. It operates on the dense Ising formulation rather than the Potts formulation used by the other implementations. The CPU implementations are not naive Monte Carlo. They exploit the uniform edge weights of this problem, and because updates are sequential, partition population counts are maintained incrementally rather than recomputed at each node, reducing memory access relative to solving the full Ising graph.
 
On CPU, runtime differences between strict and MFC are modest since updates run sequentially regardless. On FPGA, MFC enables parallel updates on the sparse graph. The right panel of Fig.~\ref{fig:hardware_advantage} plots the probability of reaching the KaFFPaE reference cut against reported solver time, using the timing definitions given in Sec. \ref{sec:tts_protocol}. The FPGA curve has a steeper slope than the CPU curves at low and intermediate success probabilities, so its time advantage is largest in the budget-constrained regime and narrows as all solvers approach certainty. The FPGA reaches the 50\% success threshold more than an order of magnitude faster than the CPU implementations and more than two orders of magnitude faster than the Tabu baseline.

KaHIP (KaFFPaE) provides a high-quality reference cut, not a runtime baseline. KaHIP is a domain-specific multilevel partitioner, whereas the techniques introduced here, p-dits and mean-field constraints, target the broader class of general-purpose Potts machines now being developed across hardware substrates, including CMOS SPAD arrays~\cite{whitehead2023cmos}, coupled CMOS ring oscillators~\cite{gonul2024potts,cheng2025oscillator}, coherent optical loops~\cite{inoue2022coherent}, and p-dit circuits~\cite{duffee2025pdit}. The Tabu solver is the natural runtime comparison; it is a heuristic combinatorial solver operating on the dense Ising encoding that constrained problems would otherwise demand, so our comparison directly isolates the speedup obtained by restoring sparsity.
 
The FPGA runtime scales linearly with sweep count and matches predicted clock-cycle timing. The FPGA curve in Fig.~\ref{fig:hardware_advantage} reports kernel time without the approximately ten seconds of host-device communication overhead incurred by the current prototype, where the beta schedule is uploaded at runtime. The speedup statements above are based on the 50\% success points of the log-time regressions shown in Fig.~\ref{fig:hardware_advantage}, using the timing protocol described in Sec. \ref{sec:tts_protocol}. Including the current prototype communication overhead adds an approximately constant offset over the sweep range where the 50\% crossing occurs; however, the 800-step annealing schedule requires only two stored values per step, \(\beta\) and \(\beta\lambda\). Thus, an implementation with the schedule stored on chip is expected to achieve timing close to the reported FPGA kernel-time curve, apart from small setup and final-readout overheads. 

A strict-constraint FPGA implementation was not pursued because exact balancing would reintroduce dense interactions, requiring sequential updates and negating hardware parallelism~\cite{Lucas_2014,Sajeeb_2025}.
 
These results show that mean-field constraints make parallel hardware acceleration practical for globally constrained problems. By restoring sparsity without sacrificing solution quality, MFC allows p-dit solvers to recover the performance traditionally associated with sparse Ising machines.

\section{Discussion}
\label{sec:discussion}

Mean-field constraints provide an effective mechanism for restoring sparsity, but they are inherently approximate. MFC does not preserve the exact constrained energy landscape and therefore does not guarantee correct Boltzmann sampling~\cite{jain2018meanfieldapproximationinformationinequalities}. As a result, mean-field constraints are best suited for optimization rather than applications requiring faithful equilibrium sampling.

Because constraint feedback is applied at the sweep level rather than instantaneously, the dynamics form a delayed feedback loop. Large gain values ($\lambda$) or fast update rates ($\alpha$) can cause oscillatory behavior as the system overcorrects~\cite{Astrom_Murray_2010}. Stabilizing the dynamics requires careful hyperparameter selection and may limit how tightly constraints can be enforced. Problems requiring strict satisfaction of hard constraints with very tight tolerances may not be well suited to a purely mean-field formulation. A two-stage strategy may offer the best of both approaches: use MFC to quickly reach a feasible region, then refine with strict enforcement.

This work introduces two complementary mechanisms for reducing constraint-induced density. p-dits handle local constraints exactly by embedding them into the node state space, preserving the energy landscape while eliminating infeasible configurations. Mean-field constraints handle global constraints approximately through shared feedback signals, restoring sparsity at the cost of exact fidelity. Together, these techniques allow constrained problems to be decomposed according to constraint structure and addressed with the appropriate strategy.

Restoring sparsity has a direct impact on hardware efficiency. The FPGA results demonstrate that MFC, combined with p-dit dynamics, achieves a speedup of more than one order of magnitude over the CPU implementations in FPGA-kernel versus CPU-solver-loop time comparisons, while maintaining solution quality. Although demonstrated on an FPGA, the principles apply broadly to GPUs, ASICs, and emerging probabilistic substrates.

Several directions follow naturally: more sophisticated feedback controllers (adaptive or learned), systems with multiple interacting constraints combining strict and relaxed enforcement, and integration with other sparsification techniques to expand the class of problems addressable on probabilistic hardware. 

While constraints add density to graphs, not all underlying unconstrained graphs are inherently sparse. Therefore, another promising direction is generalizing MFC techniques to relax specific graph nodes that have high edge density as a means to sparsify base graphs. 

\section{Methods}
\label{sec:methods}

\subsection{Detailed Balance}
\label{subsec:detailed_balance}

To establish that the p-dit update rule with nearest-neighbor candidate selection samples from the correct stationary distribution, we verify that the resulting Markov chain satisfies detailed balance. Detailed balance requires that, for any two states $i$ and $j$,
\[
\pi_i P_{ij} = \pi_j P_{ji},
\]
where $\pi_i$ denotes the stationary probability of state $i$ and $P_{ij}$ is the transition probability from $i$ to $j$.

Under the nearest-neighbor candidate selection scheme, transitions are allowed only between adjacent states on the Potts ring. The transition probability can therefore be written as
\[
P_{ij} =
\begin{cases}
A_{ij}/2, & j = (i \pm 1)\ \mathrm{mod}\ Q, \\
0, & \text{otherwise},
\end{cases}
\]
where the factor of $1/2$ accounts for uniform selection between the two neighboring candidates and $A_{ij}$ is the acceptance probability defined in Eq.~\eqref{eq:pdit_prob},
\[
A_{ij} = \sigma\!\left(\beta E(i \rightarrow j)\right).
\]
To ensure that the stationary distribution is equivalent to the Boltzmann distribution, the candidate selection must be symmetric and ergodic. We therefore use uniform selection between the two neighboring states, which satisfies both conditions.

Substituting this transition probability into the detailed balance condition yields
\[
\frac{\pi_j}{\pi_i} = \frac{A_{ij}}{A_{ji}} = \exp\!\left(-\beta\left[E_j - E_i\right]\right),
\]
which implies that the stationary distribution satisfies
\[
\pi_i \propto \exp(-\beta E_i).
\]
Since this relation holds for any pair of neighboring states, it extends recursively to all states in the ring. The p-dit dynamics therefore satisfy detailed balance and converge to the Boltzmann distribution.

While this work focuses on nearest-neighbor candidate selection for hardware efficiency, the above argument generalizes to a broader class of proposal mechanisms. Any candidate selection scheme that is probabilistic, ergodic, symmetric, and excludes self-transitions will satisfy detailed balance when combined with the acceptance rule in Eq.~\eqref{eq:pdit_prob}. Different proposal mechanisms trade off mixing speed against implementation complexity.

\subsection{p-dit Validation}
\label{subsec:pdit_val}

To validate the statistical correctness of the p-dit update rule, we measured the finite-size critical temperatures $T_c(L)$ of the two-dimensional ferromagnetic $q$-state Potts model using p-dit Monte Carlo dynamics. The model is defined by the Hamiltonian
\[
H = -\sum_{\langle i,j \rangle} \delta(s_i,s_j),
\]
on an $L \times L$ square lattice with nearest-neighbor interactions. Open boundary conditions were used in all simulations.

To obtain $T_c(L)$, the system was simulated across inverse temperatures $\beta \in [0.1,2.0]$ in steps of $0.01$. For each $(L,q,\beta)$ configuration, ten independent trials were performed, each consisting of $10^6$ p-dit updates per spin, following standard Monte Carlo sampling procedures~\cite{newman1999monte}.
 
The magnetization was measured as a function of $\beta$, and the critical point for each trial was identified as the location of the maximum derivative of the magnetization curve. The maximum derivative was approximated by first smoothing the magnetization curve with a moving average over an 11-point window. The smoothed curve was then differentiated using central finite differences over the $\beta = 0.01$ grid, taking the mean of the forward and backward differences at each point. These values were averaged across trials to obtain the reported $T_c(L)$, which were subsequently used for the finite-size scaling fits in Fig.~\ref{fig:pdit}c.

The critical temperatures in the thermodynamic limit are known exactly from Baxter’s solution of the two-dimensional Potts model~\cite{baxter1973pottsa,baxter1982exact},
\[
T_c(\infty) = \frac{1}{\ln(1+\sqrt{q})}.
\]
The corresponding correlation-length critical exponents for the $q = 2,3,4$ universality classes are also known exactly~\cite{wureview1982, Nienhuis1984}, $\nu = 1$, $5/6$, $2/3$, respectively.

Using these known quantities, the measured critical temperatures were fitted to the finite-size scaling form
\[
T_c(L) \approx T_c(\infty) + a L^{-1/\nu},
\]
with $a$ as the single fitting parameter~\cite{PhysRevLett.28.1516,PhysRevB.30.322,PhysRevB.30.1477}. The resulting coefficients were $a = -1.420$, $-1.662$, $-3.081$ for $q = 2,3,4$ respectively.

\subsection{CPU Simulations}

All CPU-based simulations were implemented in C++ using a Monte Carlo framework that directly simulates p-dit dynamics. Simulations were executed on an AMD Ryzen 7940HS CPU, with strict and mean-field constraint enforcement differing only in the constraint update mechanism.

\subsection{FPGA Implementation}
\label{sec:methods_fpga}

The hybrid Ising machine was implemented on an Alveo U250 FPGA and operated at a clock frequency of 100~MHz. The probabilistic subsystem implements p-dit-based Potts dynamics, while a classical feedback controller computes and applies the mean-field constraint. A host CPU communicates the simulated annealing schedule over PCIe. All experiments reported in Sec.~\ref{sec:hardware} use three-state p-dits corresponding to the three target partitions.

The probabilistic subsystem and feedback controller are sequentially activated. A full update of both systems takes 37 clock cycles; 2 cycles are allocated to the probabilistic subsystem to update both color group. The remaining 35 cycles are required for a pipelined implementation of the feedback controller meant to maximize the inter-system clock frequency. A majority of this time (32 cycles) is produced by a batched population count; the remaining 3 cycles are attributed to calculating and distributing the new MFC bias.

To reduce computational complexity and maximize throughput, energy differences are evaluated directly rather than computing absolute energies. For each node $i$, the local energy change associated with a proposed transition $s_i \rightarrow c_i$ is decomposed into a cut term and a balancing term,
\[
\beta E_i(s_i \rightarrow c_i) = \beta E_{\text{min},i}(s_i \rightarrow c_i) + \beta E_{\text{bal},i}(s_i \rightarrow c_i).
\]

The cut contribution is computed as
\[
\beta E_{\text{min},i}(s_i \rightarrow c_i) = \sum_{j \in \Omega_i} f(s_i,c_i,s_j),
\]
where $\Omega_i$ denotes the neighborhood of node $i$ and
\[
f(s_i,c_i,s_j) =
\begin{cases}
-\beta, & s_i = s_j, \\
\beta, & c_i = s_j, \\
0, & \text{otherwise}.
\end{cases}
\]
$\beta$ values are expressed as 10-bit fixed point (Q6.3).

The balancing contribution is computed using filtered population counts,
\[
\beta E_{\text{bal},i}(s_i \rightarrow c_i) =
\beta\lambda \hat{N}_{s_i} - \beta\lambda \hat{N}_{c_i} - \beta\lambda,
\]
where the filtered population vector is updated according to
\[
\hat{\mathbf{N}} = \alpha \mathbf{N} + (1-\alpha)\hat{\mathbf{N}},
\]
and $\mathbf{N}$ denotes the instantaneous population count for each state. The product $\beta\lambda$ combines the inverse temperature with the constraint weight $\lambda$ and is precomputed by the CPU once per beta step, then passed as a single value to the FPGA. This eliminates a per-p-dit multiplication.
 
This formulation allows both the cut objective and the mean-field balancing constraint to be evaluated using local information and a small number of global registers, avoiding dense pairwise interactions and enabling efficient parallel execution.

Table~\ref{tab:fpga_utilization} reports the resource utilization and energy per sweep for this implementation. Most of the logic is dedicated to the parallel p-dit update blocks described above. The population-count and low-pass-filter logic occupy resources comparable to a single p-dit update block, but use the DSP slices for the multiplication required to scale the filtered population counts. All reported BRAM is instantiated by the PCIe wrapper used to transfer data to and from the main module and is not accessed by the update kernel during runtime.

\begin{table}[t]
\centering
\caption{FPGA resource utilization and energy for the implementation. Resource-utilization values are post-place-and-route. The energy per sweep is computed from the active power and the 37-cycle sweep time at 100 MHz.}
\begin{tabular}{lccc}
\hline
Resource / metric & Value & Available & Utilization \\
\hline
LUTs  & 304\,363 & 1\,728\,000 & 17.61\% \\
FFs   & 123\,046 & 3\,456\,000 & 3.56\% \\
BRAM  & 81 & 2\,688 & 3.01\% \\
DSPs  & 3 & 12\,288 & 0.02\% \\
\hline
Active power (Est.) & 10.719 W & -- & -- \\
Energy per sweep & 3.97 $\mu$J & -- & -- \\
\hline
\end{tabular}
\label{tab:fpga_utilization}
\end{table}

\subsection{Annealing and Hyperparameters}
\label{sec:anneal}

This section reports the annealing schedules and hyperparameter values used for the data presented in the manuscript. Two beta schedules were used. The data in Fig.~\ref{fig:mfc}d and Fig.~\ref{fig:hardware_advantage} were collected using a linear temperature drop $\beta_n = ((T_f - T_0)\frac{n}{N} + T_0)^{-1}$, where $T_0$ and $T_f$ are the starting and ending temperatures, $n$ is the current sweep index, and $N$ is the total number of sweeps. $T_0$ was set to 3 for the 4elt graph and 8 for the cube graph, and $T_f = 0.01$ for both. To reduce the communication overhead with the FPGA implementation, the described beta schedule is approximated with a staircase function with 800 steps, where each step contains the same number of Monte Carlo sweeps. The data in Fig.~\ref{fig:mfc}e used an exponential schedule $\beta_n = \beta_0 \left(\beta_f/\beta_0\right)^{n/N}$ with $\beta_0 = 0.33$ and $\beta_f = 10$.

The MFC hyperparameters $\lambda$ and $\alpha$ were chosen as follows. For the 4 and 5 partition cases on 4elt, $\lambda = 20/15606$ and $\alpha = 0.1$. For the 32 partition case on 4elt, $\lambda$ was increased to $450/15606$ to accommodate the tighter balancing required by smaller per-partition counts, with $\alpha$ kept at $0.1$. The cube graph used $\lambda = 0.1$ and $\alpha = 0.125$; this $\alpha$ was chosen so that the FPGA low-pass filter could be implemented with bit-shift operations.

D-Wave's Tabu solver~\cite{dwave-tabu} requires an Ising encoding which reintroduces the one-hot exclusivity constraint and an additional hyperparameter $\gamma$ controlling its weight. For the Tabu baseline, the best results were obtained with $\gamma = 10.0$ and $\lambda = 0.05$.

\subsection{Time-to-solution protocol}
\label{sec:tts_protocol}

The time-to-solution data in Fig.~\ref{fig:hardware_advantage} were obtained from independent runs at each plotted runtime. Each FPGA point is averaged from 1000 independent runs and each CPU strict, CPU MFC, and Tabu point is averaged from 100 independent runs. A run is counted as successful only if the final partition reaches the KaFFPaE reference cut \(C=177\) while also satisfying the 1\% maximum-imbalance constraint. To the best of our knowledge, this cut is the minimizing balanced cut for this instance and is unique up to the permutation of partition labels.

For the CPU p-dit solvers, the reported time is the solver-loop runtime from the start of annealing to the end of annealing; one-time problem construction and file I/O are excluded. For the Tabu baseline, the reported time is the timeout value supplied to the heuristic, with one-time Ising construction excluded. For the FPGA curve in Fig.~\ref{fig:hardware_advantage}, the reported time is the measured kernel time and excludes host-device communication and runtime schedule upload.

The current host-driven FPGA prototype also incurs communication overhead because the annealing schedule is uploaded at runtime. Including this overhead, the measured end-to-end prototype times for \(10^3, 10^4, 10^5, 10^6, 10^7,\) and \(10^8\) sweeps are 9.928, 9.922, 9.917, 10.260, 13.610, and 46.370 s, respectively. These end-to-end prototype times are reported to characterize the current setup, but they are not used in the log-time regressions in Fig.~\ref{fig:hardware_advantage}, which compare solver-loop scaling.

The 50\% crossings labeled in Fig.~\ref{fig:hardware_advantage} are obtained from the dotted log-time regressions. For each solver, the empirical success probabilities \(p_{\mathrm{succ}}\) are fitted by least squares to
\begin{equation}
    p_{\mathrm{succ}}(t)=m\ln(t)+b,
\end{equation}
using the plotted time points. The reported time-to-solution is then
\begin{equation}
    t_{50}=\exp\left(\frac{0.5-b}{m}\right).
\end{equation}
The regression is used only to interpolate the 50\% crossing shown in the figure.

\section*{Data availability}
The data that support the findings of this article are not publicly available. The data are available from the authors upon reasonable request.

\section*{Code availability}
The computer code used in this study is available from the corresponding author upon reasonable request.\\ 

\section*{Author contributions}
KC-C and KYC conceived the study and wrote the manuscript. KYC supervised the study. KC-C conducted experiments regarding p-dit verification, MFC scaling on FPGA and CPU, and developed FPGA structure. KL introduced the minimum-cut formulation for p-dits and supervised CPU testing. KJ conducted strict constraint and MFC testing and benchmarking for CPU study. All authors reviewed the manuscript.\\

\section*{Competing interests}
The authors declare no competing interests. 

\section*{Acknowledgments}

This material is based upon work supported by, or in part by, the Army Research Laboratory under grant number W911NF-24-1-0228. Thanks to Christian Duffee and Pedram Khalili Amiri for fruitful discussions regarding ASIC implementations.

\onecolumngrid
\clearpage

\setcounter{section}{0}
\setcounter{figure}{0}
\setcounter{table}{0}
\setcounter{equation}{0}
\renewcommand{\thesection}{\Alph{section}}
\renewcommand{\thefigure}{S\arabic{figure}}
\renewcommand{\thetable}{S\arabic{table}}
\renewcommand{\theequation}{S\arabic{equation}}

\begin{center}
    {\Large \textbf{Supplementary Information}}\\[0.5em]
\end{center}

\section{Sparsity}
\label{supp:sparsity}

Both p-dits and mean-field constraints work by absorbing the density that constraints introduce into a Hamiltonian, leaving the underlying interaction graph sparse. This section gives a general overview of where each method applies and reports the sparsity gained for the graphs used in the main text.

In principle, a single p-dit can represent a variable with arbitrarily many states, including states that would otherwise require many coupled Ising spins. However, the nearest-neighbor two-state comparison used in our p-dit update rule mixes more slowly per update than the equivalent dense Ising encoding, so p-dits are best used to absorb local constraints rather than to replace Ising spins outright. For the graph partitioning problem, the relevant local constraint is the exclusivity (or one-hot) constraint on each vertex. Encoding this constraint with Ising spins introduces additional nodes and edges that scale with the number of partitions as
\begin{equation*}
    \text{Nodes}_{\text{Ising}} = k\cdot \text{Nodes}_{\text{Graph}} \qquad \text{Edges}_{\text{Ising}} = \frac{1}{2}\left(k^2 \cdot \text{Nodes}_\text{Graph} + k\cdot \text{Nodes}_\text{Graph}^2 -2k\cdot \text{Nodes}_\text{Graph}\right)
\end{equation*}
where $k$ is the number of partitions and the base graph is assumed to have all-to-all connections due to the balancing constraint. The p-dit implementation removes the exclusivity constraint entirely and is independent of the number of partitions.

Mean-field constraints are an approximate algorithm that absorbs global constraints into a bias signal. For graph partitioning, the global balancing constraint creates an all-to-all coupled system when enforced strictly. When replaced with the MFC approximation, the all-to-all balancing edges are removed entirely and the resulting graph density depends only on the underlying graph being partitioned.

The benefit of using these techniques is clear from a sparsity perspective. Table~\ref{tab:sparsity} reports the node and edge counts for the graphs used in the main text under three implementations: a fully Ising encoding with all constraints enforced, a Potts encoding with only the global balancing constraint enforced strictly, and a Potts encoding with the balancing constraint replaced by MFC. For 4elt at $k=32$, combining the Potts encoding with MFC reduces the edge count by roughly five orders of magnitude relative to the Ising formulation, and the Potts+MFC row matches the baseline graph density exactly. Note that the Potts node and edge counts for 4elt are the same across all $k$ values because the Potts encoding uses one variable per graph vertex regardless of the number of partitions; only the per-node state space changes with $k$.

\begin{table*}[h]
  \centering
  \caption{Node and edge counts for the 4elt and Cube benchmarks under each encoding. The Ising encoding uses $k$ binary spins per graph vertex plus all-to-all balancing edges. The Potts encoding uses one $k$-state variable per vertex with all-to-all balancing edges. The Potts+MFC encoding replaces the balancing edges with a global bias signal, recovering the density of the baseline graph.}
  \label{tab:sparsity}
  \renewcommand{\arraystretch}{1.2}
  \setlength{\tabcolsep}{3.5pt}
  \footnotesize
  \resizebox{\textwidth}{!}{%
  \begin{tabular}{l c c c c c c c c}
    \hline \hline
    \textbf{Implementation} & \multicolumn{6}{c}{\textbf{4elt}} & \multicolumn{2}{c}{\textbf{Cube}}\\
     & \multicolumn{2}{c}{$k=4$} & \multicolumn{2}{c}{$k=5$} & \multicolumn{2}{c}{$k=32$} & \multicolumn{2}{c}{$k=3$} \\
    & Nodes & Edges & Nodes & Edges & Nodes & Edges & Nodes & Edges \\
    \hline
    Baseline & 15606 & 45878 & 15606 & 45878 & 15606 & 45878 & 1000 & 2700 \\
    Ising & 62424 & 487156896 & 78030 & 608985135 & 499392 & 3904246656 & 3000 & 1501500 \\
    Potts & 15606 & 121765815 & 15606 & 121765815 & 15606 & 121765815 & 1000 & 499500 \\
    +MFC & 15606 & 45878 & 15606 & 45878 & 15606 & 45878 & 1000 & 2700 \\
    \hline \hline
  \end{tabular}
  }
\end{table*}

\section{Stability}
\label{supp:hyperparameters}

Because the mean-field constraint reacts to the instantaneous error signal with a one-sweep delay, it forms a lagged feedback loop that can oscillate if the response is too aggressive. To suppress these oscillations, the error signal needs to be filtered. In this paper we use a simple first-order IIR low-pass filter.

The graph partitioning problem has two hyperparameters relevant to stability. The first is $\lambda$, which scales the magnitude of the controller response. Setting $\lambda$ too large produces near-deterministic oscillations as the controller overcorrects. The second is $\alpha$, the filter coefficient ($\alpha=1$ means no filtering), which sets how aggressively high-frequency noise is damped. We note that $\lambda$ also appears in the original problem formulation, where it weights the importance of the balancing objective relative to the cut objective.

Finding the exact stability region is intractable in general since it depends on the energy landscape of the underlying graph, so some tuning of $\lambda$ and $\alpha$ is required. A useful heuristic comes from linearizing the feedback loop around a fixed point. The loop gain is proportional to $\lambda$ and the filter time constant is approximately $\tau \approx -1/\ln(1-\alpha)$ sweeps. Oscillations arise when the gain-bandwidth product exceeds a threshold set by the system's response time, which gives an intuition for why larger $\lambda$ requires smaller $\alpha$ (longer filter memory) to remain stable. A full nonlinear stability analysis is left to future work.

\begin{figure}[t]
    \centering
    \includegraphics[width=1\linewidth]{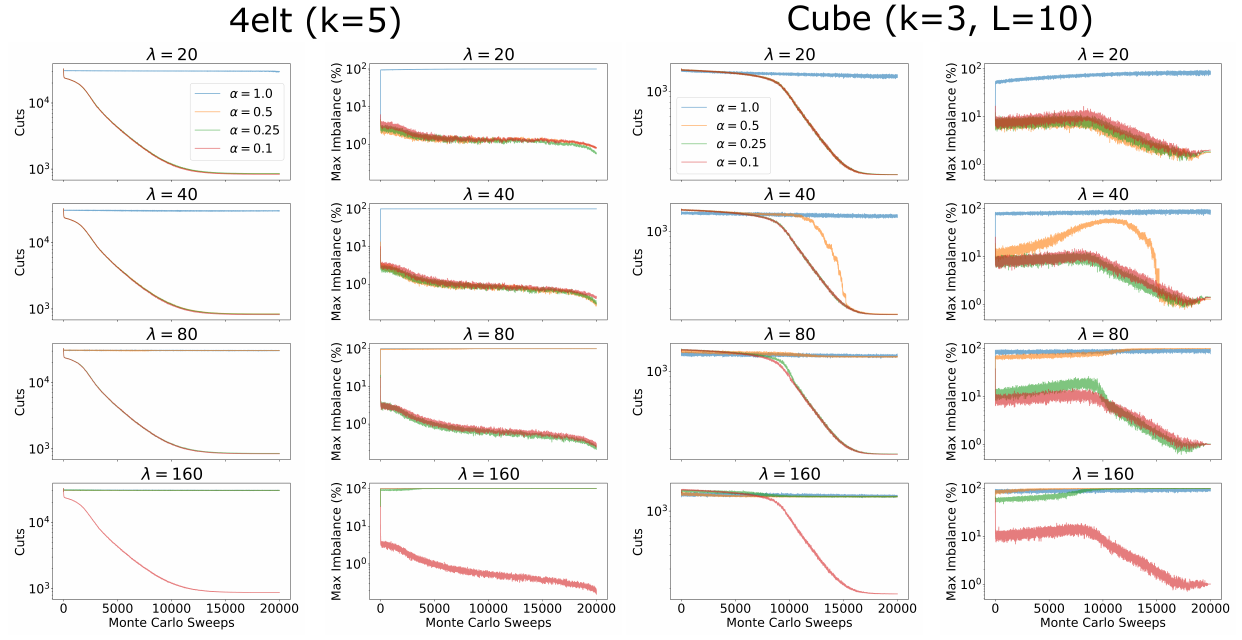}
    \caption{Hyperparameter sweep on 4elt ($k=5$, left two columns) and Cube ($k=3$, $L=10$, right two columns). Each row corresponds to a fixed gain $\lambda$, and each curve within a row corresponds to a different filter coefficient $\alpha$. Columns show cut size and maximum partition imbalance over Monte Carlo sweeps. For each $\lambda$, there is a minimum $\alpha$ below which the system converges, and over-filtering only weakly affects convergence speed.}
    \label{fig:hyperparameters}
\end{figure}

Figure~\ref{fig:hyperparameters} sweeps both parameters on the 4elt and Cube benchmarks to show how they interact in practice. As the response magnitude increases, oscillations get worse and the low-pass filter has to remove more high-frequency content to compensate. $\lambda$ needs to be tuned for both the MFC and strict solvers based on the required imbalance. For a given $\lambda$, there is a minimum $\alpha$ required for convergence, but the speed of convergence is only weakly affected by over-filtering, so erring on the side of more filtering is forgiving.

The filter choice here was made for simplicity. More complex filters could be used depending on requirements, and time-varying or adaptive filters might provide better convergence at larger gains.
 
\section{Mean-Field Constraints on GPU}

\begin{figure}[h]
    \centering
    \includegraphics[width=1\linewidth]{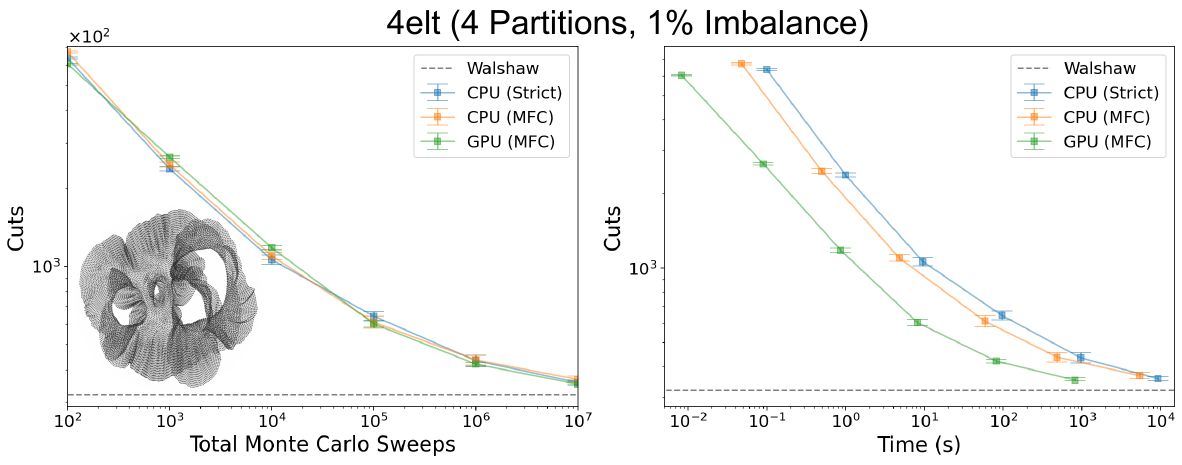}
    \caption{GPU acceleration on the 4elt benchmark (4 partitions, 1\% imbalance). Left: cuts vs. Monte Carlo sweeps; the GPU MFC implementation tracks the CPU implementations across the full sweep range, as expected since the algorithms are identical. Right: cuts vs. wall-clock time; the GPU implementation reaches comparable solution quality nearly an order of magnitude faster than either CPU implementation.}
    \label{fig:gpu}
\end{figure}

Mean-field constraints were designed to accelerate parallel hardware solvers. The main text demonstrated this on a highly parallelizable nearest-neighbor cube graph using an FPGA. The 4elt benchmark could not be implemented on the FPGA due to capacity limitations, so this section reports an alternative implementation on a GPU to show that the MFC sparsity restoration translates to general-purpose parallel hardware as well.

The solver was implemented as a single CUDA kernel written in C++ and run on an NVIDIA Tesla T4 GPU via Google Colab. The kernel performs the bulk of the computation, running p-dit updates on an entire color group at a time. Each update consists of computing an energy delta, stochastically updating the state, and atomically updating the partition counts. The full GPU solver is coordinated by a Python manager which handles beta scheduling, MFC bias updates, and kernel launches. For each Monte Carlo sweep, one kernel is launched per color group.

Figure~\ref{fig:gpu} extends the 4-partition data from Fig.~3e of the main text with the GPU solver included. The GPU dynamics closely match the CPU MFC dynamics as expected, since both implementations use the same algorithm and no additional approximations are introduced. The right panel plots cuts against wall-clock time, where the GPU is nearly an order of magnitude faster than the CPU at every iteration count even with this naive implementation.

We note that this GPU implementation, like the FPGA implementation in the main text, updates the MFC bias synchronously between sweeps. This works cleanly here because the p-dit framework lets us pause the probabilistic subsystem, compute the new bias on the classical side, and then resume the next sweep. Some Ising solvers cannot be paused mid-evolution in this way, which would require an asynchronous MFC update where the bias is computed and broadcast while the probabilistic subsystem continues to evolve. Extending MFC to handle that asynchronous setting is left to future work.

\end{document}